\documentclass[10pt,journal,compsoc]{IEEEtran}

\ifCLASSOPTIONcompsoc
  \usepackage[nocompress]{cite}
\else
  \usepackage{cite}
\fi
\usepackage[pdftex]{graphicx}

\usepackage{amsmath}
\usepackage{amsthm}
\usepackage{graphicx}
\usepackage{enumerate}
\usepackage{algorithm}
\usepackage{algorithmicx}
\usepackage{algpseudocode}
\usepackage{amssymb}
\usepackage{bm}
\usepackage{mathrsfs}
\usepackage{caption}
\usepackage{subcaption}
\usepackage{color}
\usepackage{multirow}
\usepackage{lscape}
\usepackage{url}
\usepackage{cuted}
\usepackage{microtype}

\newcommand{\sun}[1]{\textcolor{red}{sun: {#1}}}

\algdef{SE}[DOWHILE]{Do}{doWhile}{\algorithmicdo}[1]{\algorithmicwhile\ #1}

\date{}
\newtheorem{definition}{\textbf{Definition}}

\newcommand*{\rom}[1]{\expandafter\@slowromancap\romannumeral #1@}
\newcommand{\RN}[1]{\textup{\uppercase\expandafter{\romannumeral#1}}}
\newcommand{\tabincell}[2]{\begin{tabular}{@{}#1@{}}#2\end{tabular}}

\begin{document}
\title{Monitoring-based Differential Privacy Mechanism Against Query-Flooding Parameter Duplication Attack}

\author{
Haonan~Yan,
Xiaoguang~Li,
Hui~Li,
Jiamin~Li,
Wenhai~Sun
and~Fenghua~Li
\IEEEcompsocitemizethanks{
\IEEEcompsocthanksitem 
H. Yan, X. Li, H. Li and J. Li are with the State Key Laboratory of Integrated Services Networks, School of Cyber Engineering, Xidian University, Xi’an, China.
\protect\\
H. Li is the corresponding author. E-mail: lihui@mail.xidian.edu.cn.
\IEEEcompsocthanksitem 
X. Li and W. Sun are with the department of computer and information technology, Purdue University, West Lafayette, USA.
\IEEEcompsocthanksitem 
F. Li is with the State Key Laboratory of Integrated Services Networks, Institute of Information Engineering and Chinese Academic of Sciences, Beijing, China 
}
}

\IEEEtitleabstractindextext{%
\begin{abstract}
Public intelligent services enabled by machine learning algorithms are vulnerable to model extraction attacks that can steal confidential information of the learning models through public queries. Though there are some protection options such as differential privacy (DP) and monitoring, which are considered promising techniques to mitigate this attack, we still find that the vulnerability persists. In this paper, we propose an adaptive \textit{query-flooding parameter duplication} (QPD) attack. The adversary can infer the model information with black-box access and no prior knowledge of any model parameters or training data via QPD. We also develop a defense strategy using DP called monitoring-based DP (MDP) against this new attack. In MDP, we first propose a novel real-time \textit{model extraction status assessment} scheme called \textit{Monitor} to evaluate the situation of the model. Then, we design a method to guide the differential privacy budget allocation called APBA adaptively. Finally, all DP-based defenses with MDP could dynamically adjust the amount of noise added in the model response according to the result from \textit{Monitor} and effectively defends the QPD attack. Furthermore, we thoroughly evaluate and compare the QPD attack and MDP defense performance on real-world models with DP and monitoring protection.

\end{abstract}

}

\maketitle
\IEEEdisplaynontitleabstractindextext
\IEEEpeerreviewmaketitle

\IEEEraisesectionheading{
\section{Introduction} \label{Introduction}
}

\IEEEPARstart{A}{t} present, many popular machine learning (ML) models, including logistical regression and neural networks, are deployed in the secure cloud to offer services via cloud-based prediction APIs. This commercial form, known as ML-as-a-service (MLaaS), allows model developers to charge end users on a pay-per-query basis, such as Amazon, Google, Microsoft \cite{ Amazon, Google, Speech, Azure}.
This service brings convenience while also amplifying the privacy issues of the model, because the training datasets of deployed models may be proprietary and confidential, and models may also be privacy-sensitive as they could leak information about data \cite{tramer2016stealing, ateniese2015hacking}.
Recent works show that the adversary could be a user of an MLaaS platform and exploit \textit{model extraction attacks} \cite{tramer2016stealing, 7943475} to infer the internal states of machine learning models and steal the deployed model by a series of intelligent quires, thus avoiding future query payments. 
Besides, knowing the parameters of the model obtained by extraction can be used to evade attacks, especially when the model is utilized in security applications such as malware or spam classification \cite{dwork2008differential, papernot2017practical, fredrikson2015model}.

Some traditional schemes resisting model extraction attack fall into two types: monitoring-based defense \cite{kesarwani2018model, juuti2019prada} and perturbation-based defense  \cite{dwork2008differential, dwork2014algorithmic, soria2017individual, 8835087}.
Monitoring-based methods
alert abnormal behavior if a predefined threshold has been reached. But such approaches are not reliable in practice as the threshold needs to be carefully determined to defeat sophisticated attempts.
At the same time, we find that these solutions are limited by their designs and have some limitations in terms of response range and speed.
As for perturbation-based solutions, they are proposed to constrain the information returned to the users, such as differential privacy (DP).
DP as a straightforward idea has been recognized as an effective privacy-preserving technique against model extraction attack, 
since it can provide a rigorous mathematical definition of privacy and produce controllable noises for obfuscation with measurable utility loss.
We find that the only existing DP solution called BDPL has been proposed to resist the model extraction attack so far, and it works well \cite{zheng2019bdpl}.

\begin{figure}[!t]
	\centering
	\includegraphics[width=\linewidth] {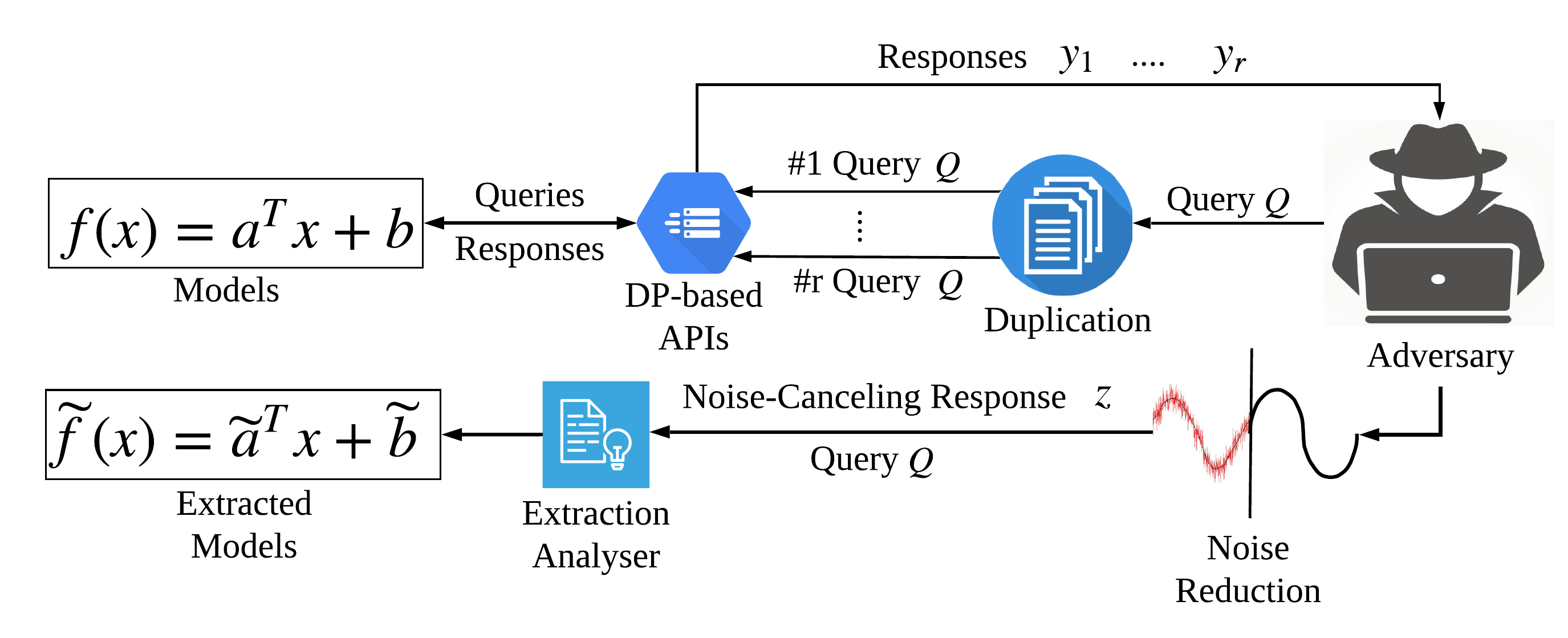}
	\caption{Illustration of QPD attack.}
	\label{contrast}
\end{figure}

Although DP-based mechanisms resist the model extraction attack plausibly, we still find there are some flaws.
The existing DP protection mechanisms utilize a fixed privacy budget to generate the identically-distributed noise for protection, which results in noise that can be attenuated in practice.
Due to the current use of deterministic perturbation methods, it is not difficult for the adversary to infer the model by sending a small number of queries and statistically estimating the true query result to solve equations.
In this work, we propose a novel combination of ordinary model extraction and multi-time query attack, called \textit{query-flooding parameter duplication} (QPD) attack.
For a $n$-dimensional machine learning model $f(\bm{x})$, the adversary first constructs a query group $\bm{Q}$ that contains at least $n + 1$ linearly independent queries. The group $\bm{Q}$ is then adaptively duplicated before being sent to the DP-protected service APIs. On receiving sufficient result vectors $\{\bm{y}_1$...$\bm{y}_r\}$, the adversary can recover the true result vector $\bm{z}$ of $\bm{Q}$ by reducing the DP noises based on the law of large numbers \cite{li2019optimal}. In the end, an inferred model $\widetilde{f}(\bm{x})$ can be derived by analyzing $\bm{Q}$ and $\bm{z}$.
The process of the QPD attack is shown in Figure \ref{contrast}.

As a result, we implement the QPD attack on existing logistic regression and neural network models with protection mechanisms.
Accordingly, we demonstrate that these defenses are vulnerable to the QPD attack.
For example, we optimize our QPD attack to adaptively exploit model extraction to boost attack in the face of BDPL protection.
The model with BDPL protection cannot withstand this new attack, and the attack effect is very obvious as it does not design a privacy budget setting strategy.
Under the hood, the existing DP-based mechanism is vulnerable when facing the hybrid attack, or the privacy budgets are excessively large.
This is also the underlying principle of the QPD attack.
Besides, QPD attack disguises as the normal query, so directly using those common defense strategies can severely restrict the availability of the service. 
Therefore, a sophisticated privacy budget setting strategy crucially dominates whether a DP-based mechanism can resist this new type of attack. 
Unfortunately, to our best knowledge, there is no way to set privacy budgets for DP-based mechanisms against the QPD attack yet. This has also been an open problem concerning the DP implementation in practice \cite{tramer2016stealing}.

Based on this situation, we believe that limiting the amount of information obtained by attackers and increasing the cost of getting information can solve this problem.
More specifically, it is advantageous to slow down the QPD attack and forecast the extraction status of future time intervals to generate alerts in advance.
To this end, 
we first design a new real-time \textit{model extraction status assessment} method called \textit{Monitor} from the perspective of model knowledge. 
By quantifying the loss of privacy of training data from such attacks, \textit{Monitor} only needs to access queries to estimate the model information gained by users relative to the training datasets of the deployed model.
Compared with previous work, \textit{Monitor} has a significant improvement in accuracy and efficiency without any additional deployment and can be applied to all types of models. 
Then, we propose an \textit{adaptive differential privacy budget allocation} (APBA) method to guide the privacy budget allocation for all DP-based defenses. This kind of improved scheme called \textit{monitoring-based differential privacy} (MDP) has the ability to generate noise dynamically.
Just as its name implies, MDP adaptively adds noise into the model responses according to the result of model extraction status from \textit{Monitor} and effectively resists the QPD attack.
It is also straightforward to apply MDP to other existing DP-based defense to respond effectively to attacks.
We finally do the theoretical analysis to demonstrate the practicality of the MDP defense and the experiment result shows its effectiveness.

Our contribution can be summarized as follows:
\begin{itemize}
\item \textbf{Adaptive QPD attack.} 
We propose a query-flooding parameter duplication attack that can be exploited to infer protected models effectively. 
We also optimize the attack process by adjusting the sending data based on the real-time feedback, which adaptively performs model extraction attacks. With optimization, QPD uses as few queries as possible and ensuring attack effectiveness.
	
\item \textbf{Real-time model extraction status assessment.}
We develop the information-theoretic metric that can quantify the loss of privacy of training data from such attacks. 
This new assessment method is independent of specific extraction attack techniques and can be applied to all models. 
It also has significant advantages in terms of evaluating accuracy and speed without any additional deployment.
	
\item \textbf{Adaptive differential privacy budget allocation and Dynamic MDP defense.} 
We design a differential privacy budget allocation method (APBA) to guide DP-based defenses to allocate the privacy budget adaptively.
In this way, these improved defenses can effectively resist QPD attacks, which are also called MDP defenses.
According to the precise feedback of model extraction status, MDP defenses are able to adjust the privacy budget dynamically and flexibly adds noise to maximize the usability of the model while effectively ensuring model privacy.

\item \textbf{Comprehensive evaluation and open source for reproducibility.} 
We implement the QPD attack and develop a prototype system for MDP defense evaluation. The result shows the validity of the QPD attack and the MDP defense. 
We are publishing the relevant source code to help the community better understand the attack and defense we proposed.
\end{itemize}

\noindent
\textbf{Roadmap.}
The rest of the paper is organized as follows: 
Section \ref{RelatedW} reviews related prior research;
Section \ref{attackmodel} introduces the principle and algorithm of QPD attack;
Section \ref{Defense} elaborates on the design of \textit{Monitor} and the process of applying APBA into MDP defense;
Section \ref{Experiment} presents our measurement study and experiment results;
Section \ref{discussion} discusses the limitations of our current design and potential future research and Section \ref{Conclusion} concludes the paper.

\section{Related Work}\label{RelatedW}

\subsection{Model Extraction Attack}
\textit{Tram$\grave{e}$r et al.} \cite{tramer2016stealing} propose the model extraction attack on machine learning models (e.g. logistic regression, decision tree, SVM and neural network). They also demonstrate the existence and the harm of the model extraction attack in the experiments.
Later on, many researchers focus on making this attack more effective and efficient.
\textit{Papernot et al.} \cite{papernot2017practical} propose a Jacobin-based synthetic data augmentation technique to train a synthetic DNN model by model extraction attack at first. Then this model is used to construct adversarial samples to mislead the deployed model. They also prove that this attack is practical for adversaries with no knowledge about the model. 
\textit{Juuti et al.} \cite{juuti2019prada} propose a new type of model extraction attack using a novel approach to generate synthetic queries and optimize training hyperparameters.
The extraction attack performs effectively on DNN models.
\textit{Shi et al.} \cite{7943475} discover an exploratory model extraction attack by deep learning, which can steal functionalities of Naive Bayes models and SVM classifiers with high accuracy.
\textit{Duddu et al.} \cite{duddu2018stealing} infer the depth of the neural network by timing side channels and use reinforcement learning to accelerate the extraction process.
\textit{Wang et al.} \cite{8418595} propose a hyperparameter stealing attack for both non-kernel models and kernel models with the strong assumption that the knowledge of training data and algorithm is available. Their method highlights the need for new defenses against the stealing attacks.
However, it is worth noting that all the mentioned attacks target non-DP protected systems.
In this work, we propose the QPD attack, the first of its kind, to extract model information even with strong DP obfuscation.


\subsection{Resistance to Model Extraction Attack}
To defend against model extraction attacks, 
the initial idea is to reduce the amount of information leaked to the adversarial user by modifying the results from the deployed model \cite{tramer2016stealing,lee2018defending}.
Then, \textit{Quiring et al.} \cite{8406619} propose to exploit the closeness-to-the-boundary concept in digital watermarking to mitigate the attack against decision trees.
However, \textit{Juuti et al.} \cite{juuti2019prada} prove that this type of defense method is ineffective. They also illustrate the digital watermark schemes have a high false alarm rate and do not apply to high dimensional input spaces.
Subsequently, \textit{Kesarwani et al.} \cite{kesarwani2018model} presents a model extraction warning method that quantifies the extraction status of models by continually checking the API queries and responses.
Their approach is over-dependent on the decision tree model as the proxy model, which is not suitable when the deployed model is too complex. It also assumes that the distribution of benign inputs needs to be the same as the training datasets.
Besides warning, \textit{Zheng et al.} \cite{zheng2019bdpl} propose boundary differentially private layer (BDPL)  for binary classifier. To our best knowledge, this is the only DP solution to mitigate model extraction, and the defense effect of this mechanism is very significant. Unfortunately, they do not deliberate the privacy budget setting, and their experiments show that the extraction rate could be close to 90\% when using a large number of queries.

\section{Query-flooding Parameter Duplication Attack} \label{attackmodel}
In this section, we introduce the query-flooding parameter duplication attack. The QPD attack can be exploited to extract the coefficients of models protected by state-of-the-art protection mechanisms.

\subsection{Setup}

\noindent\textbf{Basic Assumptions.}
To better summarize the attack model, we put forward four basic assumptions about the capabilities of the adversary as below:
\begin{itemize}
	\item the adversary has access to the APIs of the target machine learning models as a regular user or a group of colluding users;
	\item the adversary knows the model type, e.g., logistic regression or neural network model;
	\item the adversary knows the dimensionality of the target model by accessing the public APIs;
	\item the adversary can adapt query generation by observing the system responses to optimize the attack efficacy.
\end{itemize}

And under the basic assumptions, the attack method could be designed later.

\noindent\textbf{Symbol Description.}
We define a slice of symbols when constructing the method, as shown in Table \ref{notations}.
\begin{table}[h]
	\footnotesize
	\centering
	\caption{NOMENCLATURE} \label{notations}
	\begin{tabular}{|l|l|}
		\hline
		\textbf{Notation}                              & \textbf{Description}                                         \\
		\hline
		$n$                                            & \tabincell{l} { The number of dimensions of the model}       \\
		\hline
		$r$                                            & The number of duplicates in QPD attack                       \\
		\hline
		$f(\bm{x})$                                    & Original machine learning model                 \\
		\hline
		$\bm{Q}_i^k$                                   & The $k$-result matrix for $\bm{Q}_i$                         \\
		\hline
		$z^{(i)} \in \bm{z}$                           & The result corresponding to the $i$-th query $\bm{q}^{(i)}$  \\
		\hline
		$\bm{q}^{(j)} \in \bm{Q}_i$                    & The $j$-th query vector in $\bm{Q}_i$                        \\
		\hline
		$\bm{Q} \in \mathbb{R} ^ {m \times n}$         & \tabincell{l} {All received queries, where $m$ is the number \\ of queries and $n$ is the number of features} \\
		\hline
		$\bm{Q}_i \in \mathbb{R} ^ {(n + 1) \times n}$ & \tabincell{l} { The $i$-th query group (matrix) after random \\ arrangement} \\
		\hline
	\end{tabular}
\end{table}

Other instructions of symbols will be given in the text.

\subsection{Target Model} \label{Perliminaries}
Let $\mathcal{X}$ be a training dataset containing $m$ tuples ${(\bm{x}^{(1)}, y^{(1)}), (\bm{x}^{(2)}, y^{(2)}), ..., (\bm{x}^{(m)}, y^{(m)})}$. The $i$-th tuple $(\bm{x}^{(i)}, y^{(i)})$ includes $n + 1$ explanatory attributes (or ``dimensions", ``features") $x_{1}^{(i)}, x_{2}^{(i)}, ..., x_{n}^{(i)}, y^{(i)}$, where $(x_{1}^{(i)}, x_{2}^{(i)}, ..., x_{n}^{(i)})$ is the input of the regression model $f(\bm{x})$ and $y^{(i)}$ is the predicted value (or called ``label" in logistic regression) corresponding to the $\bm{x}^{(i)}$. Based on the above notations we introduce the necessary preliminary knowledge used in this work. We indiscriminately use "dimension" or "feature" instead of "attribute", and use "label" instead of "predicted value" for logistic regression.



\begin{definition}[Logistic Regression]
	An $n$-dimensional logistic regression model trained on dataset $\mathcal{X}$ is a prediction function which returns $1$ with probability $f(\bm{x}) = \frac{1}{1 + e ^{ - (\bm{a} ^ {T}\bm{x} + b)}}$, where $\bm{a} \in \mathbb{R}^n$ and $b \in \mathbb{R}$ are coefficients that minimize the cost function
	$
		J(\bm{a}, b) = -\sum\limits_{i = 1}^{m} \left[ y^{(i)} \log{f(\bm{x}^{(i)})} + (1 - y^{(i)}) \log{(1 - f(\bm{x}^{(i)}))} \right]
	$.
\end{definition}

\begin{definition}[Neural Network]
The Neural network is non-linear, and it is composed of three layers (i.e., input layer, hidden layer, and output layer). New inputs are transmitted to a hidden layer via the input layer. In the hidden layer, the input is weighted summed by weights $\bm{w}$ and directed to the activation function as the output. Finally, the output layer returns the classification result.

The training process of the neural network is similar to Logistic regression. We also need to create a loss function $J(\bm{w}) = dist\left( f(\bm{x}), y \right)$ to measure how good the neural network for the task. Given the loss function, the training process is equivalent to minimizing the loss function and find the optimal $\bm{w}$.
\end{definition}



\subsection{Attack details}\label{attackL}
For linear models, an adversary can carry out 
the QPD attack by combining the na\"ive equation-solving model extraction attack with the multi-query attack \cite{li2019optimal}. 
In a nutshell, given a linear model $f(\bm{x})$,
the adversary can create $n + 1$ linearly independent queries denoted by a query matrix $\bm{Q}$. 
Then, the adversary adaptively duplicates $r$ times of each query $\bm{q}^{(j)} \in \bm{Q}$  and gets the duplicated query matrix $\bm{Q}_d$.
Next, the adversary sends the query $\bm{Q}_d$ to the model $f(\bm{x})$ and receives the results $\mathcal{Y} \in \mathbb{R}^{(rn +r) \times 1}$. Because a query $\bm{q}^{(j)} \in \bm{Q}$ is duplicated $r$ times, thus each query corresponds to $r$ different perturbed results $\{y_i^{(j)}\}_{i = 1}^r \in \mathcal{Y}$. 
Finally, the adversary can solve the $n + 1$ coefficients ${\bm{a}} = [{a_1},{a_2}, \ldots ,{a_n}]$ and $b$ using the law of large numbers and the Cramer's rule.
The logistic regression model is essentially linear equations. 
The transfer functions of the linear neural network, are linear and the outputs of neurons can be written as a linear function of inputs. 
Here we only introduce an equation-solving attack for linear models. 
For the non-linear model, the QPD attack can invoke a shadow model attack scheme to infer useful approximations of query results used in the shadow model attack scheme \cite{shokri2017membership} and optimizes the number of attacks to make the attack more effective.


In theory, the larger the $r$ is, the more similar the extracted model is to the original model due to sufficient samples. But in practice, excessive duplication significantly slows down the attack without noticeable result improvement. In the QPD attack, we allow the adversary to find the optimal $r$ adaptively. Informally speaking, the adversary duplicates $\bm{Q}$ twice and sends it to the model at the beginning. Since the results follow the same distribution, he can use a hypothesis test to test if the results follow the Gaussian distribution or Laplace distribution. Based on the given distribution, he can calculate the $95\%$ confidence interval (CI) for the population mean of the perturbed results, i.e., the true result. Once the length of the CI is large than a defined threshold, he could double the $r$ and repeat the above steps again until the length of CI is smaller than the threshold.

\noindent
\textbf{Optimization of $r$.}
Concretely, this confidence interval estimation contains three steps. First, the adversary calculates the sample mean and sample variance and considers them as the population mean and variance. Second, given the estimated population mean and variance, the adversary uses hypothesis testing to test whether these samples are drawn from Laplace distribution or Gaussian distribution (since Laplace noise and Gaussian noise are the most popular DP mechanisms, we mainly discuss these two distributions, and the adversary can also do hypothesis test on other noise distributions). Note that there are many hypothesis testing methods to do this task \cite{woolf1957log, lilliefors1967kolmogorov, chen1998speaker}, we do not discuss this part since this content is beyond this paper. Third, once the distribution is determined, the adversary can calculate $95\%$ CI for the population mean. Formally, according to \cite{alrasheedi2012confidence}, we have the $95\%$ CI for the population mean of Laplace distribution is
\begin{align*}
\left[ \widetilde{X} - X_{\frac{1.95}{2}} \frac{\sum_{i=1}^n \left| X_i - \widetilde{X} \right|}{n}, \widetilde{X} - X_{\frac{0.05}{2}} \frac{\sum_{i=1}^n \left| X_i - \widetilde{X} \right|}{n} \right],
\end{align*}
where $\widetilde{X}$ is the median of all samples, and $X_{\alpha}$ denotes the $\alpha$ quantile of Laplace distribution. 

The $95\%$ CI for the population mean of Gaussian distribution is
\begin{align*}
\left[ \bar{X} - z_{\frac{0.05}{2}} \frac{1}{\sqrt{n}} s, \bar{X} + z_{\frac{0.05}{2}} \frac{1}{\sqrt{n}} s \right],
\end{align*}
where $\bar{X}$ is the sample mean of all samples, $s$ is the sample variance, and $z_{\alpha}$ denotes the $\alpha$ quantile of Gaussian distribution.

If the confidence interval is more significant than a threshold $\Delta$, it means that the $r$ is too small to infer an exact result. Then the adversary uses an exponential search to iteratively find the next optimal value of $r$ until the confidence interval is smaller than $\Delta$. By using the exponential search, we can see the optimal $r$ in logarithmic time.
\\

With $r$ determined, the adversary construct $\bm{Q}_d$ and get the result $\mathcal{Y}$, given which he can estimate the true result $z^{(j)}$ for each query $\bm{q}^{(j)} \in \bm{Q}$ according to the law of large numbers:

\begin{align}
	z^{(j)} = \frac{1}{r} \sum_{i = 1}^{r} y_i^{(j)}
\end{align}

Finally, he has $n + 1$ equations denoted by the augmented matrix $\bm{Q}_A$.

\begin{align*}
	 & \quad \bm{Q}_A = \left( \bm{Q} | \bm{z} \right) \\
	 & =
	\left(
	\begin{array}{cccccc|c}
			x^{(1)}_1     & \cdots & x^{(1)}_k     & \cdots & x^{(1)}_n     & 1      & z^{(1)}     \\
			\vdots        & \ddots & \vdots        & \ddots & \vdots        & \vdots & \vdots      \\
			x^{(n + 1)}_1 & \cdots & x^{(n + 1)}_k & \cdots & x^{(n + 1)}_n & 1      & z^{(n + 1)}
		\end{array}
	\right)
\end{align*}

where $x^{(j)}_k$ is the value in the $k$-th dimension of the $j$-th query $\bm{q}^{(j)}$.
Then we replace the $k$-th column of query matrix $\bm{Q}$ with the column $\bm{z}$ to construct the $k$-result matrix $\bm{Q}^k$.

\begin{align*}
	\bm{Q}^k =
	\left(
	\begin{array}{cccccc}
			x^{(1)}_1     & \cdots & z^{(1)}     & \cdots & x^{(1)}_n     & 1      \\
			\vdots        & \ddots & \vdots      & \ddots & \vdots        & \vdots \\
			x^{(n + 1)}_1 & \cdots & z^{(n + 1)} & \cdots & x^{(n + 1)}_n & 1
		\end{array}
	\right)
\end{align*}

As a result, the coefficient can be derived by the Cramer's rule:

\begin{align*}
	a_k = \frac{det(\bm{Q}^k)}{det(\bm{Q})} = \frac{\left|\bm{Q}^k\right|}{\left|\bm{Q}\right|}
	\text{,  }
	b = \frac{det(\bm{Q}^{n + 1})}{det(\bm{Q})} = \frac{\left|\bm{Q}^{n + 1}\right|}{\left|\bm{Q}\right|}
	\text{,  }
	k \leq n.
\end{align*}

The QPD attack is shown in the Algorithm \ref{QPD attack}.

\begin{algorithm}[!thbp]
	\caption{\textbf{Query-Flooding Parameter Duplication Attack (QPD)}}
\leftline{\textbf{Input:} \text{DP-protected $n$-dimensional regression model}
	\text{$f(\bm{x})$; } 
}
\leftline{\text{threshold $\Delta$}}
\leftline{\textbf{Output:} \text{Extracted $n$-dimensional regression model $\widetilde{f}(\bm{x})$}
}
	\begin{algorithmic}[1]
		
		\State Construct a query matrix $\bm{Q}$ containing $n + 1$ linearly independent queries $\bm{q}^{(j)}$
		\State $r \gets 1$ \Comment{number of duplicates}
		\Do \Comment{find the optimal $r$}
		\State $\bm{L} \gets \emptyset$, $r \gets r \times 2$
		\State $\bm{Q}_d \gets \text{Duplicate } \bm{Q} \text{ } r \text{ } \text{times}$
		\State Send $\bm{Q}_d$ to $f(\bm{x})$ and get query result $\mathcal{Y} \in \mathbb{R}^{(rn + r) \times 1}$
		\For {$j := 1$ to $n + 1$} \Comment{Calculate length of 95\% CI}
		\State Calculate sample mean $\hat{\mu}$ and variance $s$ of $r$ query results $\{ y_i^{(j)} \}_{i = 1}^r \in \mathcal{Y}$ corresponding to $\bm{q}^{(j)}$
		\State Hypothesis test if the noise follows Laplace distribution or Gaussian distribution
		\State Calculate length $l_j$ of CI for the population mean given noise distribution
		\State $\bm{L} \gets \bm{L} \cup \{ l_j \}$
		\EndFor
		\doWhile {there exists $l \in \bm{L}$ greater than $\Delta$}
		\For {$j := 1$ to $n + 1$}  \Comment{estimate true results}
		\State Calculate mean $z^{(j)}$ of $r$ query results $\{ y_i^{(j)} \}_{i = 1}^r \in \mathcal{Y}$ corresponding to $\bm{q}^{(j)}$
		\EndFor
		\State Construct column vector $\bm{z} = [z^{(1)}, z^{(2)}, ..., z^{(n + 1)}] ^ T$
		\For {$k := 1$ to $n$}   \Comment{solve coefficients}
		\State Construct $k$-result matrix $\bm{Q}^k$ by replacing the $k$-th column in $\bm{Q}$ by $\bm{z}$
		\State $a_k = \frac{\left| \bm{Q}^k \right|}{\left| \bm{Q} \right|}$
		\EndFor
		\State $b = \frac{\left| \bm{Q}^{n + 1} \right|}{\left| \bm{Q} \right|}$, $\bm{a} = [a_1, a_2, ..., a_n]$
		\State \Return $\widetilde{f}(\bm{x}) = \bm{a}^T \bm{x} + b$
	\end{algorithmic} \label{QPD attack}
\end{algorithm}

\section{Monitoring-based Differential Privacy Defense}\label{Defense}

\begin{figure}[!tbp]
	\centering
	\includegraphics[width=\linewidth] {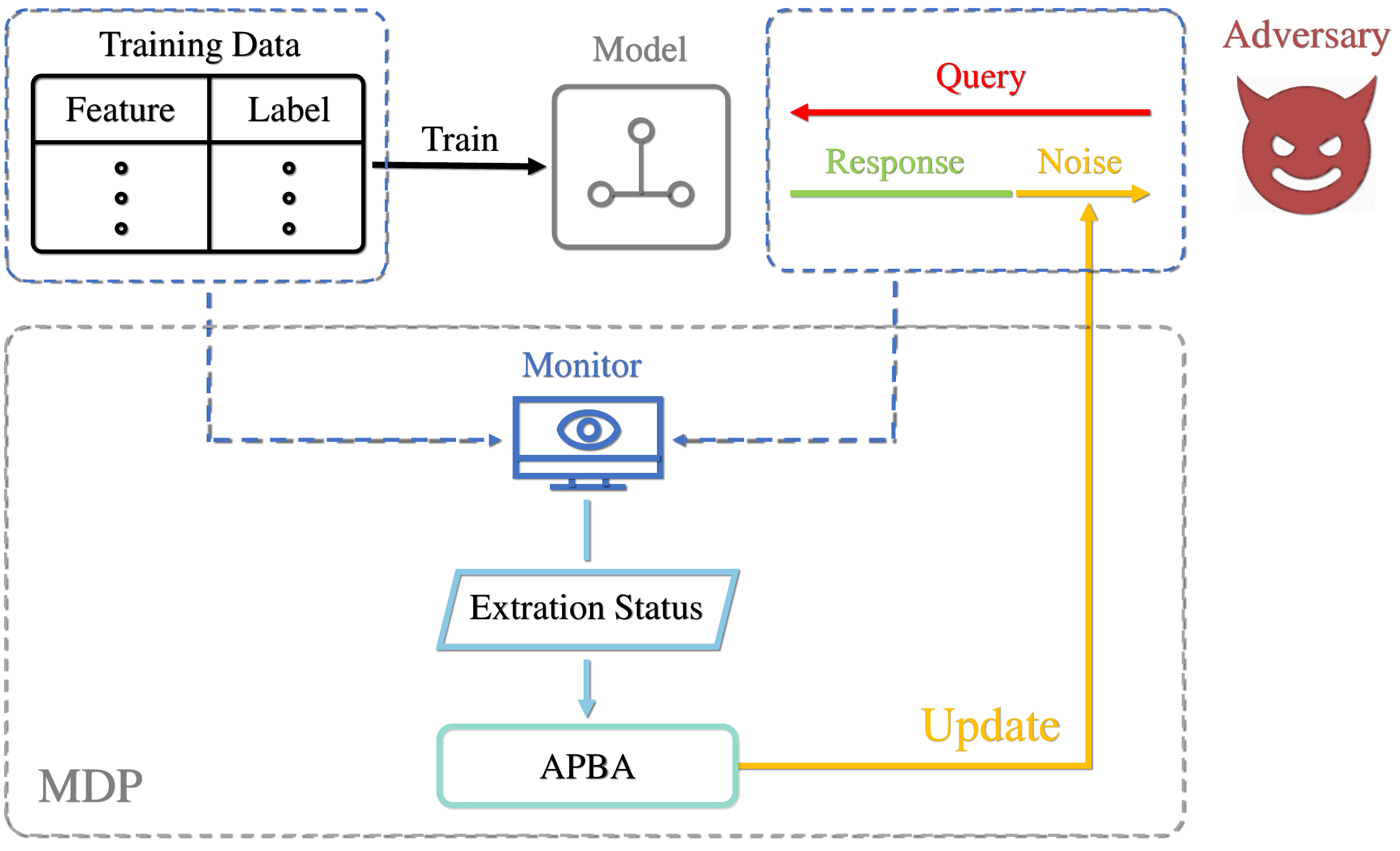}
	\caption{Illustration of MDP Defense.}
	\label{mdp}
\end{figure}

In this part, we propose a novel method that combines the advantages of monitor and differential privacy to mitigate the QPD attack. 
The core idea is as follows. 
First, we use a monitor to supervise the queries and measure the accumulated information leakage for individuals and colluding adversaries. In the meanwhile, we use a DP-based mechanism to respond to queries as to protect the model. Then the $\epsilon$ of differential privacy mechanism is adaptively updated according to the model leakage status from the \textit{Monitor}. For example, the larger the accumulated information leakage judged by Monitor, the smaller the $\epsilon$ adapted and the more noise added in the responses.
Compared with the previous single Monitor or DP protection, this new defense not only actively protects the model parameters from being stolen but also does not affect the feasibility of the model intuitively.
The structure of MDP is shown in Figure \ref{mdp}.

\subsection{Monitor: Real-time Model Extraction Status Assessment}  

Some existing schemes are available to reveal the status of model extraction, but they have some problems in terms of efficiency and feasibility. For example, surrogate decision tree models in \cite{kesarwani2018model} will bring high computational cost and low efficiency for complex models.
To this end, we propose a new approach for real-time model extraction status assessment more generally. 
This approach considers that the deployed model's knowledge comes from the training datasets, and the queries sent by the adversarial user(s) aim for the knowledge of the training datasets through the model as a pipeline in essence. 
Therefore, we develop the information-theoretic metric that can quantify the loss of privacy of training data from such attacks.
Then, we directly compare the submitted queries and model outputs against the training data to indicate the degree to which the current model has been extracted.
The problem of how to assess the status of model extraction shifts to how to assess the correlation between query matrix $\bm{Q}_A$ (see Section \ref{attackL}) and training dataset $D$.

Usually, we use the Euclidean Distance to measure the correlation between vectors, but the Euclidean Distance can not consider the difference between two variables and is sensitive to the range of variables. 
Therefore, we choose the Pearson Correlation Coefficient (PCC) \cite{benesty2009pearson} here to perform the correlation on both the queries/training samples and outputs/training labels.
Another reason for choosing PCC is that its computational complexity is $O(n)$, capable of handling high dimensional data.
It is worth mentioning that the correlation between queries is also taken into account.

In our scenario, we use $\bm{u}$ to concatenate the query $\bm{q}$ from user(s) and the response $z$ from model, where $\bm{u} = \bm{q} \| z$. 
Then $ \bm{u} $ would be appended  as a new row in $ \bm{Q}_A $.
To assess the amount of information $\bm{u} $ obtained by the $ i $-th $ \bm{q} $  of $ n $ queries and the corresponding $ i $-th response $z$ from the deployed model, we use the following equation:
\begin{equation} \label{equ2}
I_{\bm{u}} = \mathop{\rm PCC}\nolimits (D, \bm{u}) \times {\bm{I}_D} -  {\mathop{\rm PCC}\nolimits} (\bm{Q}_A, \bm{u}) \times \bm{I}_{\bm{Q}_A}
\end{equation}
where $ D $ represents the training dataset of the deployed model and $I_D$ denotes the amount of information possessed by the dataset. 
$\bm{Q}_A$ represents the matrix concatenation of total queries received from $ 1 $ to $i-1$ (including $i-1$) and their corresponding responses to user(s).
$I_{\bm{Q}_A}$ denotes the amount of information possessed by this matrix.

We further explain how to calculate the item $I_M$. We use $M$ to denote a $n \times m$ matrix, where $m$ is the number of tuples in the $M$ and $n$ is the number of features in the $M$. Here we use entropy to define $\bm{I}_M$. Formally, we have
\begin{align*}
    \bm{I}_M = H(X_1, X_2, ..., X_n),
\end{align*}
where $X_i (1 \leq i \leq n)$ is the $i$-th feature in $M$. Besides, according to the chain rule, we can also rewrite $\bm{I}_M$ as follows,
\begin{align*}
    \bm{I}_M &= H(X_1, X_2, ..., X_n) \\
    &= H(X_1) + H(X_2 | X_1) + ... + H(X_n | X_{n - 1}, ..., X_1)\\
    &= \sum_{i = 1}^{n} H(X_{i} | X_{i - 1}, ..., X_1)
\end{align*}

\noindent
\textbf{Interpretation of Equation \eqref{equ2}.}
When the user sends the $i$-th query $\bm{q}$, the model first generates the response $\bm{z}$, and concatenates $\bm{q}$ and $z$ into $\bm{u}$, then uses the $PCC$ function to calculate the correlation between $\bm{u}$ and each tuple in the training dataset to obtain the correlation vector. Finally, by doing the dot product of the correlation vector and the information quantity vector $\bm{I}_D$, we can get how much information quantity of the training dataset is contained in the query $\bm{q}$.
We also consider that queries may be related to each other, so it needs to subtract the amount of the repeated information quantity between the $\bm{u}$ and the query matrix $\bm{Q}_A$ (containing previous $i-1$ queries and corresponding responses).
It should be noted that as the number of queries increases, the amount of information obtained by per query will gradually decrease.
\\

After calculating the amount of information stolen by each query, we need to calculate the training set's total extraction degree.
The training dataset consists of data belonging to different classes. For example, suppose we have a training dataset containing labels $0$ and $1$, which means the class number is $2$.
Logically, the amount of information leaked of each class can be computed by aggregating the amount of information of all queries within the same class. 
Then, per class's extraction status can be computed as the ratio of the total amount of information leaked in each class to the total amount of all information owned by the corresponding class.
Finally, the whole model extraction status can be calculated as a weighted sum of per class extraction status, and the class probability $ {{p_k}} $ can be easily obtained from the class distribution of training set.
\begin{equation}
Extraction\_Status = \sum\limits_{k \in \bm{C}} {({p_k} \times \frac{{\sum\nolimits_{\bm{u} \in {\bm{C}_k}} {\bm{I}_{\bm{u}}} }}{{{\bm{I}_{{\bm{C}_k}}}}})} 
\end{equation}


Unlike all previous schemes, this approach does not need any additional deployment (such as proxy or surrogate models). 
It also does not assume any training data distribution but only studies samples submitted by a given client. 
Whether features are categorical or continuous, it can be used directly.
By comparing the correlation of two datasets, we can determine the extraction degree of any model without considering the model's complexity.
This explains why our method has significant advantages in both accuracy and performance.

It is worth to emphasize that what Monitor seeks is the theoretical maximum of information leakage.
Aiming for high-fidelity model extraction rather than high-accuracy extraction, we do not consider things like model architecture or hyperparameters.
The reason is that adversaries may not train an equivalent model because they may not have a mature model training experience.
However, suppose the knowledge extracted from the target model is sufficient to support an equivalent model's training. In that case, it is only a matter of time before the adversaries make full use of the knowledge.
Consequently, we maintain that model extraction status assessment should be considered in extreme cases where adversaries are at the same level as the model provider.

\subsection{APBA: Adaptive Differential Privacy Budget Allocation}\label{APBA}

Traditional DP-based mechanism against model extraction attack uses a fixed $\epsilon$ value for each query no matter how much information about the model has been leaked. 
This is a major flaw in the existing DP-based mechanism.
When the leaked information is large enough, this kind of mechanism cannot adaptively enhance protection, increasing the risk of the stolen model.

\begin{figure}[!htbp]
	\centering
	\includegraphics[scale = 0.3]{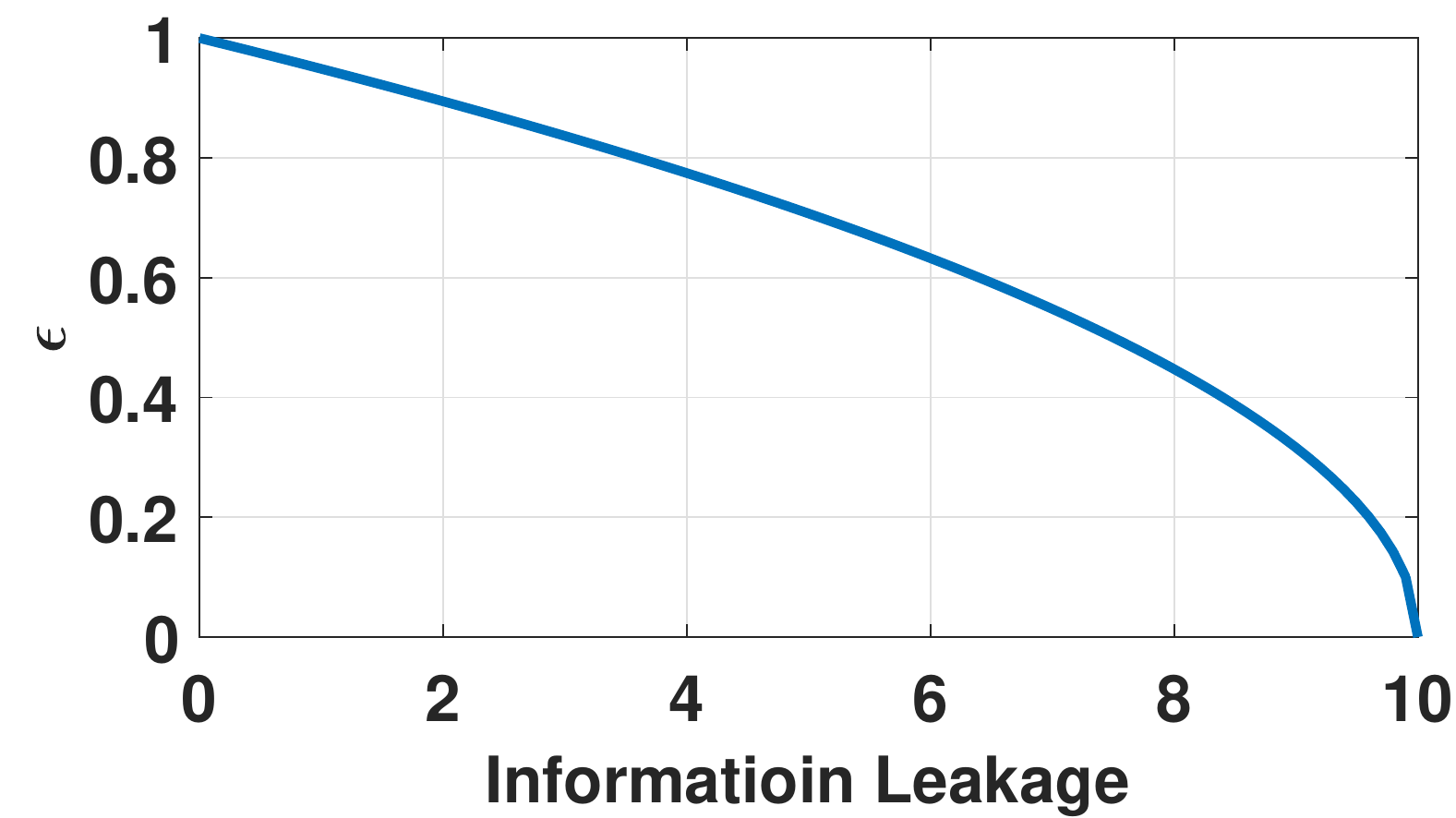}
	\caption{Example of Privacy Budget Allocation Function ($L_t = 10$, $\epsilon = \frac{20}{3}$)}
	\label{parabola}
\end{figure}

In this part, we propose a method to adaptively allocate the privacy budget $\epsilon$ every time the model responds to users' queries. 
The main idea is that we allocate the privacy budget used in the DP-based mechanism according to the information leakage degree. The more the information leakage, the smaller the allocated $\epsilon$ value. 
With this objective, we find that the parabola that opens to the left meets our requirement, and it perfectly captures the relationship between the information leakage and the privacy budget $ \epsilon $. 
Intuitively, it reduces faster and faster as the information leakage grows. 
Thus, we set the privacy budget allocation function to be $\epsilon_i = \sqrt{-p \times (L_i - L_t)}$, where $p$ is the scale parameter, $\epsilon_i$ is the $\epsilon$ value set for the $i$-th query, $L_i$ is the information leakage caused by the previous $i - 1$ queries, and $L_t$ is the threshold about the information leakage set by the model owner. 
Then we explain the scale parameter $p$. Since the privacy budget $\epsilon$ should be used up at the threshold of the information leakage, we have the following relationship between $p$ and $\epsilon$
\begin{align*}
    \int_{0}^{L_t} \sqrt{-p \times (L_i - L_t)} d L_i &= \epsilon,
\end{align*}
Then we can calculate $p = \frac{9\epsilon^2}{4 L_t^3}$.
Figure \ref{parabola} shows an example of the parabola used in our framework. Our adaptive privacy budget allocation is presented in Algorithm \ref{privacy_budget_allocation}.

\begin{algorithm}[!htbp]
	\caption{\textbf{Adaptive Privacy Budget Allocation (APBA)}}
	\leftline{\textbf{Input:} \text{Privacy budget $\epsilon$, information leakage $L_i$, information}}
	\leftline{\text{leakage threshold $L_t$}}
	\leftline{\textbf{Output:} \text{$\epsilon_i$ value for the query $\bm{q}_i$}}
	\begin{algorithmic}[1]
	    \State Calculate the remaining privacy budget $\epsilon_{re} = \epsilon - \sum_{k=1}^{i-1} \epsilon_k$
		\State Calculate the scale parameter $p = \frac{9\epsilon_{re}^2}{4 L_t^3}$
		\State Calculate the allocated privacy budget $\epsilon_i = \sqrt{-p \times (L_i - L_t)}$
	\end{algorithmic} \label{privacy_budget_allocation}
\end{algorithm}

\subsection{MDP: Monitoring-based Differential Privacy Mechanism}
Systematically, MDP needs two initial security parameters to provide differential privacy for the model: tolerable information leakage threshold $L_t$ and privacy budget $\epsilon$.
Users can customize the parameters according to the scene.
More specifically,
$L_t$ is the maximum degree of model information leakage that users can accept.
Privacy budget $\epsilon$ refers to the privacy level in differential privacy. 
$\epsilon$ controls the size of the added disturbance when the model responds to the queries.
In each subsequent query, MDP adaptively allocates the $\epsilon$ to ensure that the leakage of model information does not exceed the preset value. 
MDP works as follows:
\begin{enumerate}[\quad Step 1:]
\item Listen to the user's query.
\item Calculate the accumulated information leakage caused by previous queries of the user.
\item Allocate privacy budget $\epsilon$ to the query.
\item Return the perturbed result and listen to new queries.
\end{enumerate}
The specific process is presented in the Algorithm \ref{QPD mitigation}. 
\\

\begin{algorithm}[!tbp]
	\caption{\textbf{Monitoring-based Differential Privacy (MDP)}}
	\leftline{\textbf{Input:} \text{query $\bm{q}$, model $f(\bm{x})$, privacy budget $\epsilon$}}
	\leftline{\textbf{Output:} \text{DP-protected query results}}
	\begin{algorithmic}[1]
		\State Set the order of the query $i = 0$, accumulated information leakage $L_0 = 0$, query matrix $\bm{Q}_A = \emptyset$
		\While{Receive the query of the user $\bm{q}$ \& $\epsilon \textgreater 0$}
		\State $i = i + 1$
		\State $y = f(\bm{q})$
 		\If{i == 1}
 		\Comment{Special handling for the first query}
 		\State $L_1 = 0$
 		\Else
		\State $L_i = L_{i-1}+ I_{\bm{u}}$
 		\EndIf
 		\State $\epsilon_i = APBA(L_i, L_t, \epsilon)$ \Comment{Allocate privacy budget}
		\State $z_i = DP(y_i, \epsilon_i)$
		\State $\bm{u} = \bm{q}_i \| z_i$
		\State $I_{\bm{u}} = \mathop{\rm PCC}\nolimits (D, \bm{u}) \times {\bm{I}_D} -  {\mathop{\rm PCC}\nolimits} (\bm{Q}_A, \bm{u}) \times \bm{I}_{\bm{Q}_A}$
		\State $\epsilon = \epsilon - \epsilon_i$
		\State Append $\bm{u}$ as a new row in $\bm{Q}_A$
		\State Append $\bm{I}_{\bm{u}}$ as a new row in $\bm{I}_{\bm{Q}_A}$
        \State Reply $z$
		\EndWhile
	\end{algorithmic} \label{QPD mitigation}
\end{algorithm}


To better illustrate the applicability of MDP as we claim that MDP can be adopted to all DP-based defense schemes, here we give an instance to show how to improve the original BDPL with MDP.
For each query, BDPL is used as a basic differential privacy method to provide differential privacy protection with two parameters: boundary-sensitive zone $\Delta$ and privacy budget $\epsilon$.
$\Delta$ denotes the zone near the decision boundary in the feature space. If a query is in this zone, BDPL considers this query to have a high risk of leaking the decision boundary. Intuitively, smaller $\Delta$ incurs high model utility but high potential risk, since BDPL only perturbs a small part of queries. Given the parameters, the BDPL mechanism $\mathcal{M}$ perturbs the query answer $y \in \{0, 1\}$ by the following function.
\begin{align}
\mathcal{M}(y) = \begin{cases}
y & \text{with probability } \frac{1}{2} + \frac{\sqrt{e^{2\epsilon} - 1}}{2 + 2e^\epsilon} \\
1 - y & \text{with probability } \frac{1}{2} - \frac{\sqrt{e^{2\epsilon} - 1}}{2 + 2e^\epsilon}
\end{cases}
\end{align}
After the model owner sets the $\epsilon$ and $\Delta$, BDPL will not change the parameter in the future.
Following this method makes the adversary still successfully steal the model with the QPD attack we proposed. 

Nonetheless, an adaptive BDPL can be easily constructed with the help of MDP.
Based on the real-time results of the extracted status from the \textit{Monitor}, MDP can dynamically allocate the privacy budget. Administrators only need to use this budget to the deployed BDPL to respond to the query, and BDPL can defend against QPD attacks successfully.
This is just an example of BDPL. Other DP-based defenses can better cope with the model extraction attack after using MDP, without modifying the already deployed mechanism.
 
\section{Experimental Evaluation}\label{Experiment}
In this section, experimental performance results are presented for the QPD attack (in Sec.  \ref{attackmodel}) and MDP defense (in Sec. \ref{Defense}).
For the part of the attack, we evaluate the proposed QPD method's effectiveness to the LR and NN models.
We also simulate two protection strategies employed by model owners: \textit{monitoring-based} and \textit{perturbation-based} schemes. They are state-of-the-art in their respective directions and used to counter the QPD attack.
For the part of the defense, we mainly compare the security of our scheme MDP with the only existing DP related work BDPL under the same QPD attack (in Sec. \ref{PDdef}). 
The utility of MDP is also considered (in Sec. \ref{uti}).
Besides, model extraction status \textit{Monitor} we proposed, as the core component of MDP, is also compared with \textit{Warning} \cite{kesarwani2018model} to demonstrate performance and efficiency (in Sec. \ref{overallevaluation}).
For all the experiments, adversaries' queries are linearly independent (same as \cite{tramer2016stealing}), and the number of duplicates $r$ of QPD is determined adaptively.
With these settings, attackers can use as few queries as possible to extract the model.
All results are repeated multiple times and averaged.

\subsection{Setup}
\noindent
\textbf{Datasets.} 
We use four publicly available datasets in our experiments: 
The \textit{Email Spam} model classifies emails as spam, given the presence of certain words in its content.
\textit{Mushrooms} comes from scikit \cite{pedregosa2011scikit} and the other two come from Kaggle \cite{Kaggle}.
The four data sets are typical representatives of four orders of magnitude. 
All the datasets are split into 70\% for training and 30\% for evaluation. 
All categorical items are encoded by one-hot-encoding \cite{harris2015digital}. Using one-hot-encoding, we do not assume that the learning models understand the order among the models' dimensions. So we can improve the performance of the model by eliminating the redundant order information. 
Missing values are replaced by this attribute's mean, which is common and useful in this situation. 
The details about these datasets are shown in Table \ref{datasets}.

\begin{table}[ht]
	\centering
	\caption{Datasets}\label{datasets}
	\begin{tabular}{lrr}
		\hline
		\textbf{Dataset} & \multicolumn{1}{l}{\textbf{Instances}} & \multicolumn{1}{l}{\textbf{Dimensions}}  \\ \hline
		SocialAds       & 401   & 5     \\
		Titanic  & 1310  & 28  \\
		Email Spam & 4601 & 46   \\
		Mushrooms     & 8124  & 112  \\ 
		\hline
	\end{tabular}
\end{table}

\noindent
\textbf{Evaluation Metrics.}
We use \textit{accuracy} to measure the utility of LR and NN models. As for the extracted models, we use the ${R_{test}}$ metrics formulated by \cite{tramer2016stealing} to evaluate the performance of the extracted model as compared to the original model on the test dataset.
Therefore, $ 1-{R_{test}} $ is defined as the extraction status under test error.
It is worth mentioning that testing models on the test datasets following the same distribution as the training dataset.
The indicator can also test the extraction model on other random datasets with different distributions to estimate the fraction of the full feature space uniformly as ${R_{unif}}$.

\begin{itemize}
	\item \textit{Accuracy} measures the proportion of the correct classification results for models. It indicates the utility of the model from  perspective of users. Formally, given the model $f(\bm{x})$, the number of tuples $m$, the $i$-th tuple $\bm{x}^{(i)}$ and the corresponding label $y^{(i)}$,
	\begin{align}
	Accuracy = \frac{1}{m} \times \sum_{i = 1}^{m} d\left( f(\bm{x}^{(i)}) = y^{(i)} \right)
	\end{align}
	where $d$ is an indicator function that equals 1 if $f(\bm{x}^{(i)}) = y^{(i)}$, otherwise 0.


	\item \textit{Test error $ {R_{test}} $} measures the similarity between the extracted model and the original model. A larger $ 1- {R_{test}} $ indicates the similarity is higher, and the extraction attack is thus more effective. Formally, given the extracted model $\widetilde{f}(\bm{x})$ and the test dataset  $ {D_{test}} $,
 \begin{align}
 {R_{test}} = \frac{1}{\left| {D_{test}} \right|} \sum\limits_{i \in {D_{test}} } d\left( f(\bm{x}^{(i)})  \ne  \widetilde{f}(\bm{x}^{(i)}) \right)     
\end{align}
where $d$ is an indicator function that equals 1 if $f(\bm{x}^{(i)}) = \widetilde{f}(\bm{x}^{(i)})$, otherwise 0.

\end{itemize}

\subsection{QPD vs Monitoring-based Defense} \label{overallevaluation}
This part focuses on the effect of QPD on optimizing the number of duplication times $ r $.
The total number of QPD queries is $  r  *  (n+1) $.
The higher the dimensions $ n $ of the deployed model, the worse the results obtained by relying only on $ n + 1 $ queries ($ r = 1 $). An increase in $ r $ is needed to enhance the attack effect.
But too many queries would trigger monitoring alerts, so we optimize the $ r $ of QPD to minimize the number of queries, to ensure a stable attack effect.

\begin{figure*}[!htbp]
	\centering
	\includegraphics[width=1.0\linewidth]{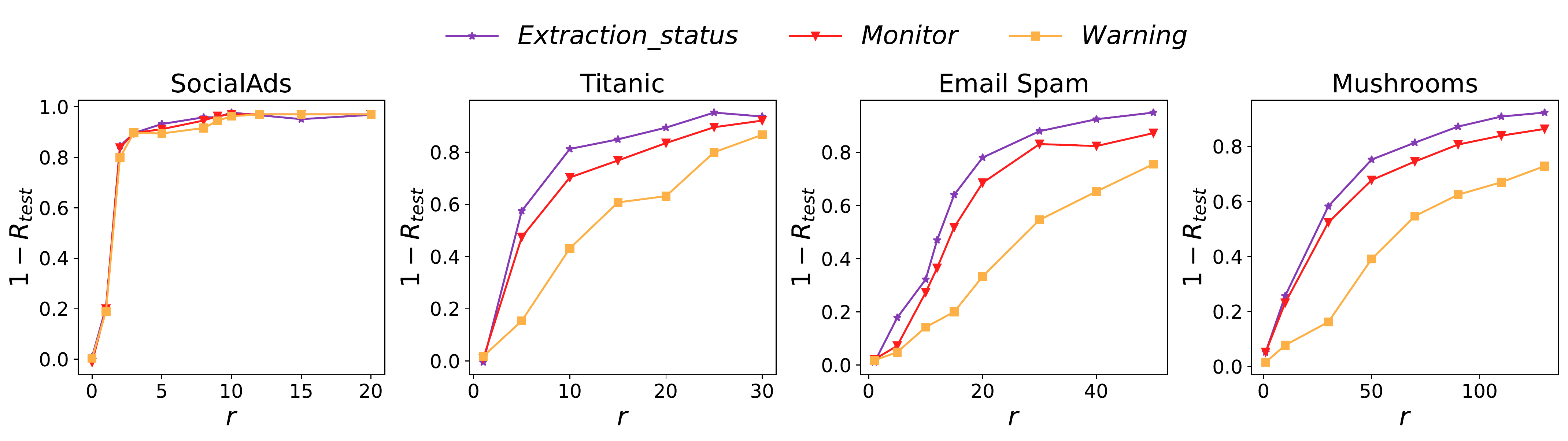}
	\caption{Comparison results of monitoring-based defenses against the QPD attack in LR models.}
	\label{QPDvsM}
\end{figure*}

\begin{figure*}[!htbp]
	\centering
	\includegraphics[width=1.0\linewidth]{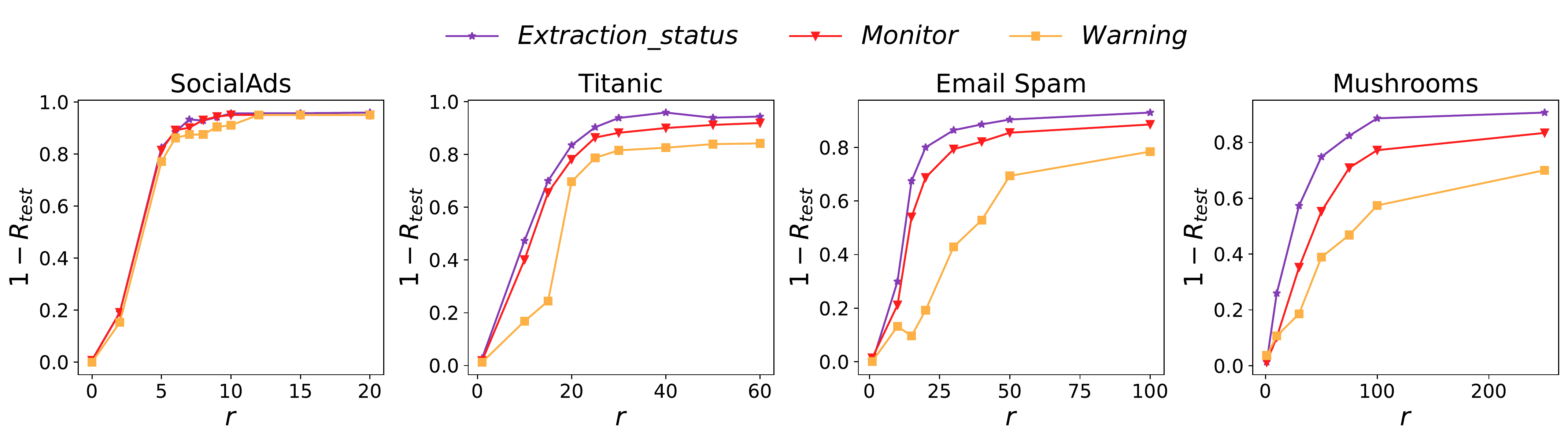}
	\caption{Comparison results of monitoring-based defenses against the QPD attack in NN models.}
	\label{QPDvsM_N}
\end{figure*}

Now there are mainly two kinds of Monitoring-based defenses schemes: \textit{Warning} of the secure threshold for the number of queries \cite{kesarwani2018model} and \textit{PRADA} for the distribution of queries \cite{juuti2019prada}.
For the \textit{PRADA} designed to protect the DNN model, it needs a sufficient number of queries to generate a query distribution to detect.
However, the QPD attack uses genuine queries and essentially relies on equation solving.
For a linear model with n-dimensional parameters, QPD completely extracts the model only by $ n + 1 $ queries ($r=1$) ideally.
The amount of the QPD query is insufficient for \textit{PARDA} to generate the distribution to compare,
enabling the attack to be implemented before the \textit{PARDA} alert is triggered,
thus this option is not considered.
We here compare the proposed \textit{Monitor} with \textit{Warning}, and the experimental results are shown in Figure \ref{QPDvsM} and \ref{QPDvsM_N}.


In experiments, the extracted model trained separately for each value of $ r $.
It can be found that there is a large gap between the two monitoring schemes. As the number of attacks ($ r $) and the complexity of the model increase, this gap is also gradually enlarged among four datasets. (SocialAds model is too simple, leading to insignificant results.)
The main reason for this gap is that \textit{Monitor} directly compares the knowledge of the deployed model (i.e., training datasets) with the knowledge of the extracted model (i.e., queries and results) to determine the extraction status in a highly efficient manner.
However, \textit{Warning} needs to convert knowledge into decision tree models or perform knowledge distillation into decision tree models in advance. This process involves loss of information. The consequence is that \textit{Warning} is delayed in judgment and limited in the types of models, compared to \textit{Monitor} we proposed.

\begin{figure}[htbp]
	\centering
	\includegraphics[width=0.9\linewidth]{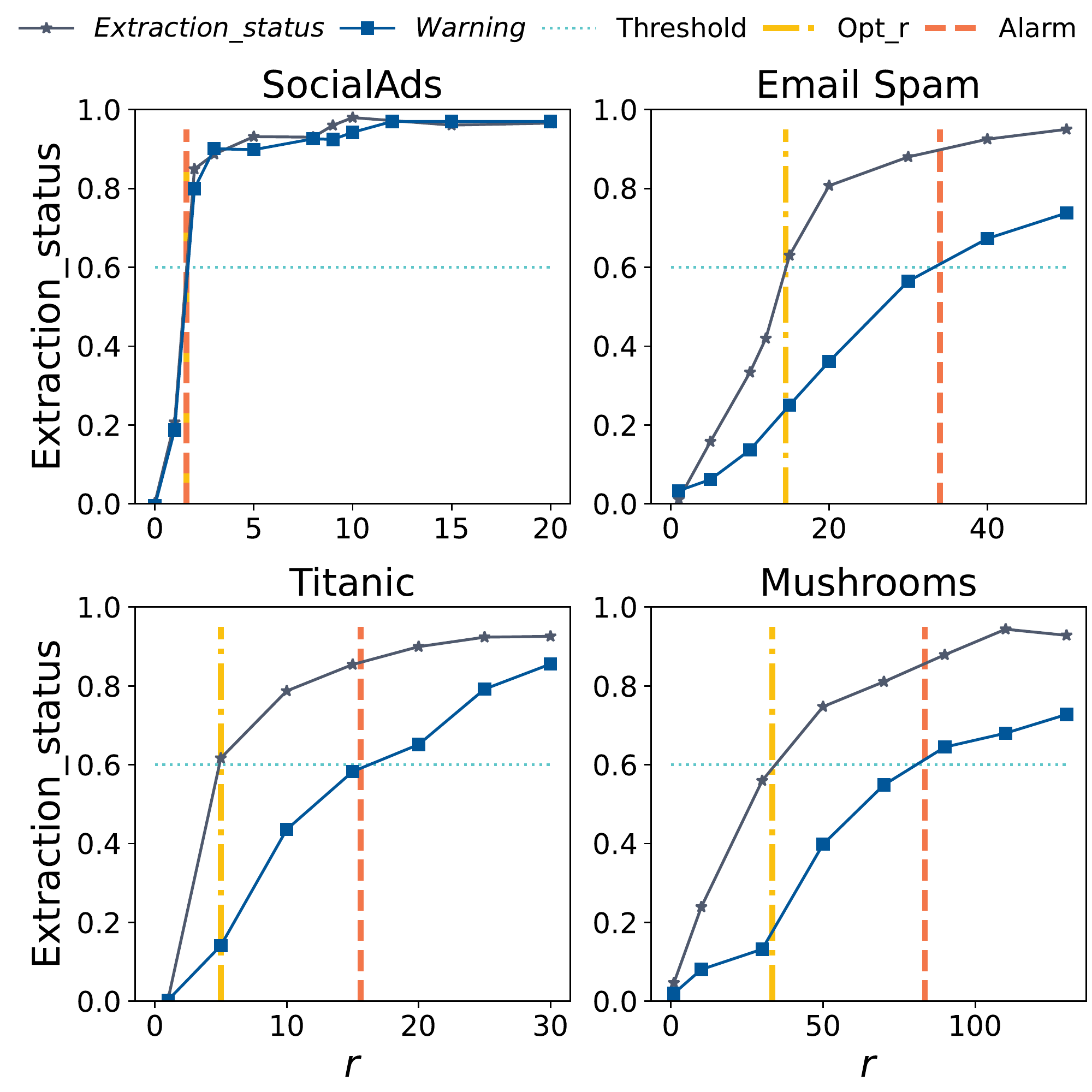}
	\caption{Impact of $r$ of QPD attack in LR models. When the model owner sets the tolerable model extraction status to 60\%, the QPD attack reaches the extraction level at $Opt\_r$, but the \textit{Warning} does not report the attack until the number of attacks reaches $Alarm$.}	
	\label{impact_of_r_on_QPD_attack}
\end{figure}

We also do experiments to show that despite the deployed \textit{Warning} mechanism, QPD attack can increase the accuracy of extracted models by at least 20\% during this delay. 
The attack can also obtain higher utility before triggering an alarm with the optimal $ r $.
The optimal $r$ is marked as \texttt{Opt\_r} and results are shown in Figure \ref{impact_of_r_on_QPD_attack}.

\noindent
\textbf{A typical example to illustrate. }
When the model owner sets the tolerable model extraction degree of the \textit{Email Spam} LR model to 60\%, \textit{Monitor} reports that the extraction degree of the model is over 60\% when $  r = 17 $, and the actual extraction status of the model is 69.90\%. Under the same experimental setting, \textit{Warning} needs $  r = 36$ to trigger the alarm. 
At this time, the model's real extraction status is 89.94\%, and \textit{Monitor} determines that the extraction level should be 82.38\%. The details can be seen in Table \ref{partialR}, and the results for other moments and datasets are presented in Figure \ref{impact_of_r_on_QPD_attack}.

\begin{table}[htbp]
	\centering
	\caption{Example of monitoring results against QPD attack. }\label{partialR}
	\begin{tabular}{cccc}
		\multicolumn{4}{c}{\textbf{Email Spam}}       \\ \hline
		& Extraction\_Status & \textit{Monitor} & \textit{Warning} \\ \hline
		r=17 & 69.90\%            & 60.74\% & 29.47\% \\
		r=36 & 89.94\%            & 82.38\% & 60.04\% \\ \hline
	\end{tabular}
\end{table}

\noindent
\textbf{Scalability analysis.}
It is worth mentioning that in the real environment, the \textit{Warning} method can limit the QPD attack by setting the threshold to the extreme case, but this causes disturbances for regular users and severely affects model usability (e.g., expanding transaction volume). In other words, when deploying such defense online, as usual, the QPD threat still exists.
Though the \textit{Monitor} detects QPD attack earlier and in a more timely manner than the \textit{Warning}, it also can not fundamentally reduce the extraction accuracy degree of models extracted via QPD.
Therefore, We add DP perturbation to slow down the extraction speed of QPD.
\textit{ Monitor} can discover the attack and trigger alerts before the extracted models are sufficiently available.

\subsection{QPD vs Perturbation-based Defense} \label{PDdef}
Perturbation-based defense schemes currently use two kinds of techniques: \textit{Differential Privacy (DP)} and \textit{Rounding Confidences (RC)}.
We evaluate the prior DP-protected LR and NN model works and try differential privacy mechanisms, i.e., DPBA \cite{8835087} and BDPL \cite{zheng2019bdpl}, in terms of utility and security.
Among the two studies, only BDPL can defend against the general model extraction attack.
Though DPBA is suitable and used for both models as it injects the noise into the cost function during the training process, it cannot resist the QPD attack since the linear property of the models persists.
Finally, we only considered BDPL in our experiments. In other words, BDPL is the only DP mechanism against the model extraction attack for classifiers.
According to the work of \cite{tramer2016stealing} rounding class probabilities further to 3 and 2 decimal places, the model extraction attack is weakened significantly.
As a result, the attack is also launched on RC-protected models to evaluate the effectiveness of QPD.

For the BDPL, we set the default $\epsilon = 1.0$ and the zone parameter $\Delta = \frac{1}{8}$ .
For the RC, we round the confidence score to 2 decimals as it provides protection while retaining a decent utility.
For the MDP, we set its $\epsilon = 1.0$ and $L_t = 100\% \times I_D$.
This setting allows the attacker to demonstrate the complete attack details more clearly.
%

\begin{figure*}[!thbp]
	\centering
	\includegraphics[width=1.0\linewidth]{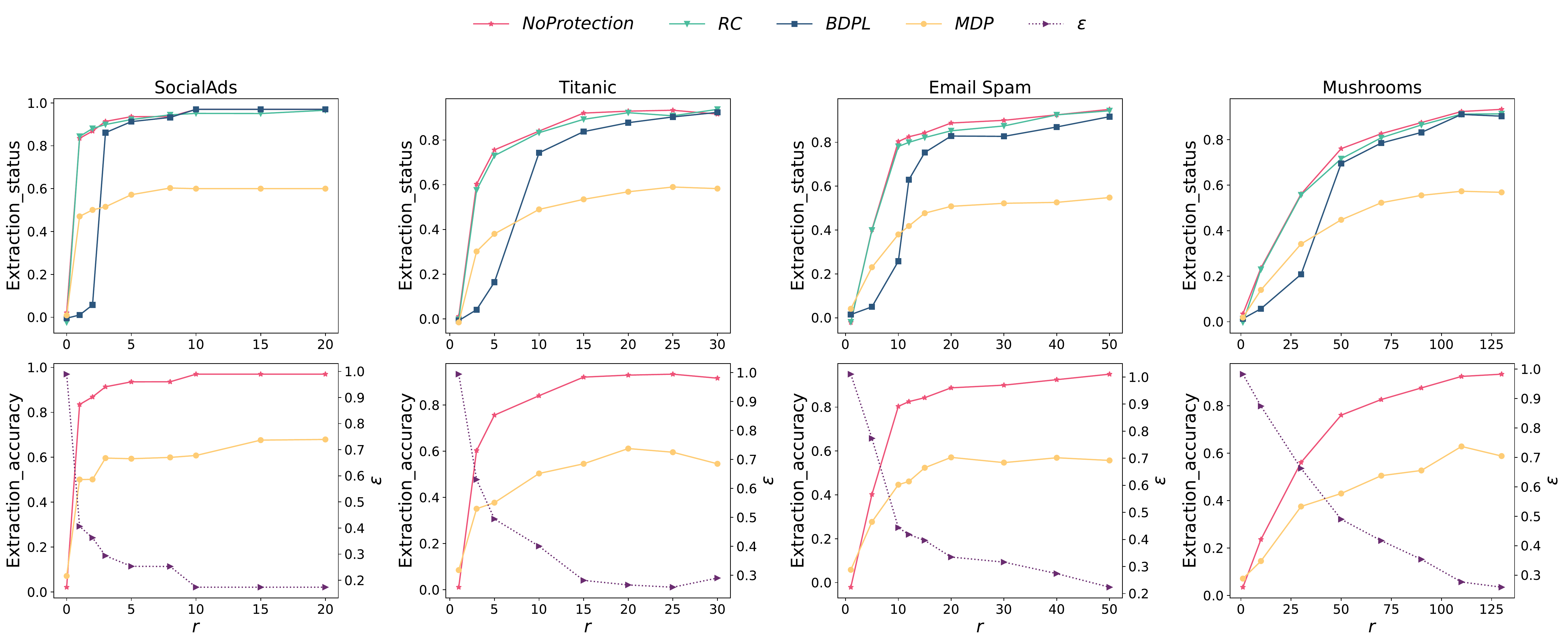}
	\caption{Comparison results of perturbation-based defenses against the QPD attack in LR models.}
	\label{LogisAttack}
\end{figure*}

\begin{figure*}[!thbp]	
	\centering
	\includegraphics[width=1.0\linewidth]{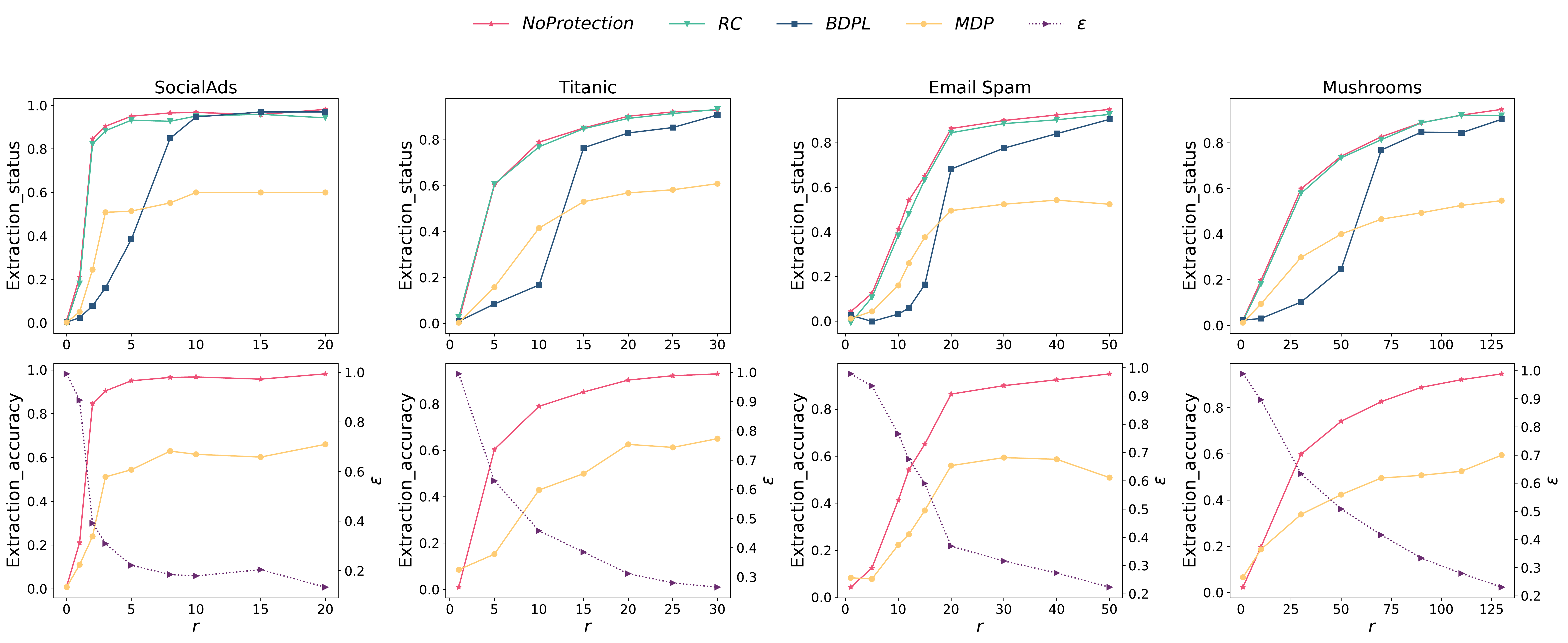}
	\caption{Comparison results of perturbation-based defenses against the QPD attack in NN models.}
	\label{NNATT}
\end{figure*}

We plot the $1 - {R_{test}} $ and $Extraction\_accuracy$ of extracted model as the functions of the duplication times $r$.
Higher $ 1-{R_{test}} $ indicates higher fidelity of the extracted model. They both are used to represent the effect of the QPD attack.
In this work, we let $r$ automatically adjust to the optimal number of duplicates.
Figure \ref{LogisAttack} and \ref{NNATT} show the protection effect for the LR and NN models.

\noindent\textbf{QPD attack effect.}
From Figure \ref{LogisAttack} and \ref{NNATT}, 
the $ 1-{R_{test}} $ and $ Accuracy $ of the extracted models by QPD remains over 80\% on all tested datasets, which is very close to the unprotected case. This exhibits that the QPD provides excellent attack effects.
Besides, according to the figures, the general trend of $ 1-{R_{test}} $ on protected models increases as the number of queries $r$ gradually grows. 
The small gap between the $ 1-{R_{test}} $ on NoProtecion, RC-protected and BDPL-protected model indicates the stable efficiency of the QPD attack, which also proves that these previous protection schemes are not very effective against QPD attack.
The bottom four subfigures in Figure \ref{LogisAttack} and \ref{NNATT} also show the accuracy of the QPD attack without any defense. 

\noindent\textbf{Defenses effect.}
Among all datasets, the protection effect of BDPL is slightly overall better than that of RC but also very limited.
The root cause is that RC only discards some decimals, which is weak protection. . 
For the BDPL, when QPD initially attacks the model, its protection is very active because of the massive noise.
But once the number of attacks increases, the defensive effect quickly wears off and dramatically degrades with larger $r$. 
This is because the same and fixed privacy budget is always used for all the queries.
To mitigate the QPD attack, we need to generate uncorrelated noises.

As a consequence, MDP dynamically adjusts the noise according to the real-time model extraction status.
Due to usability considerations, MDP adds minimal noise initially, so the degree of model extraction rises faster. 
Correspondingly, starting with a higher privacy budget allocation and a faster decline in $\epsilon$ is consistent with the design trend of APBA in Section \ref{APBA}.
However, once the extraction degree reaches criticality, the noise will steeply increase to limit the attack.
Finally, MDP completely mitigates the model extraction attack. When the deployed models face a large number of QPD attacks, MDP keeps the extraction accuracy between 50\% and 60\% and reports the abnormal queries to the administrator simultaneously.
The bottom four subfigures in Figure \ref{LogisAttack} and \ref{NNATT} show that MDP adaptively determines the privacy budget $\epsilon$ based on the query results and reduces the attack accuracy.

\subsection{Utility of MDP} \label{uti}
\begin{figure}[!ht]	
	\centering
	\includegraphics[width=0.9\linewidth]{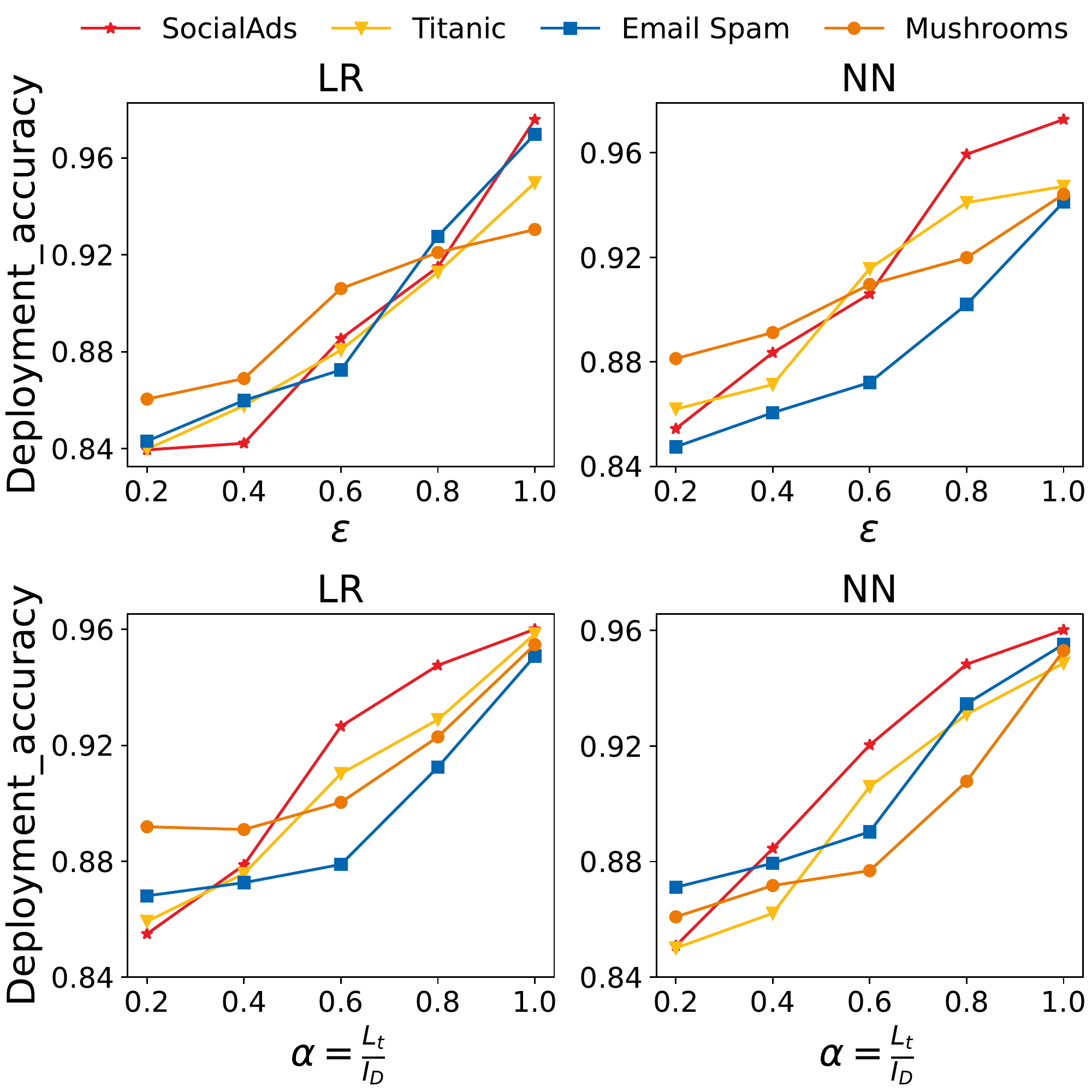}
	\caption{Impact of $\epsilon$ of MDP on its utility. }
	\label{impact_of_epsilon_on_QPD_attack}
\end{figure}

While MDP works well for model privacy protection, adding noise to the response inevitably brings interference to the model's output and affects the utility of the model. In this part, we evaluate our MDP from the perspectives of varying privacy budget $\epsilon$ and different information leakage threshold $L_t$. 
Note that $ I_D $ represents the amount of information in the training data set $ D $ of the deployed model. We denote the $L_t$ as $\alpha \times I_D$ ($\alpha \in [0, 1]$) and plot the deployed model accuracy as the function of $\alpha$.
We set the default $\epsilon = 1.0$ and the default $\alpha = 100\%$, and keep the number of query to be $20k$ to stabilize the results. The results are shown in Figure \ref{impact_of_epsilon_on_QPD_attack}.

As shown in the top two subfigures in Figure \ref{impact_of_epsilon_on_QPD_attack}, we vary $\epsilon$ to evaluate the performance. Our MDP achieves satisfactory accuracy both on logistic regression (LR) and neural network (NN). In particular, the accuracy increase from about 0.84 to about 0.96 on LR and NN as $\epsilon$ increases. 
Besides, it is also logical to show the real models' decrease in accuracy as $ r $ increases under the MDP defense.

As shown in the bottom two subfigures in Figure \ref{impact_of_epsilon_on_QPD_attack}, we analyze the performance with varying $\alpha$. The model accuracy also grows as $\alpha$ increases, i.e., from 0.84 to 0.96, approximately both on LR and NN. This is because larger $\alpha$ allows more information about the dataset to be accessed. According to our adaptive differential privacy budget allocation (APBA) algorithm, this makes $\epsilon$ value decreases more slowly and improve model accuracy.



\section{Discussion} \label{discussion}
We discuss the potential influence and limitations of the proposed QPD attack and MDP protection.
Simultaneously, we provide insights into the extended application of our methods.
\\

\noindent\textbf{Influence.}
The QPD attack is straightforward yet devastating and can successfully steal machine learning models 
if intuitively applying differential privacy or monitoring mechanism for protection. 
Similar vulnerabilities may exist in other models, such as the polynomial regression model. 
Suppose adversaries know the dimension of the polynomial. In that case, they can recover the design matrix and convert the polynomial regression into a linear system to get the coefficients via QPD, because these models can be written and solved as a system of linear equations.
For non-linear models, the shadow model attack used by QPD is completely unavoidable because it is consistent with a regular query.
Unfortunately, to our best knowledge, no mechanisms are defending against the QPD attack with these settings currently.

In response to the QPD attack, we design the MDP defensive solution. 
The core sniffer \textit{Monitor} uses model knowledge for extraction status evaluation. By quantifying the loss of privacy of training data from such attacks, it can be applied to not only LR and NN but also all machine learning models.
Relying on the accurate assessment of \textit{Monitor}, MDP can flexibly adjust the amount of added noise, which maximizes the model utility while ensuring the privacy of the model.
\\

\noindent\textbf{Limitation.}
QPD attack also has some noticeable limitations in practice.
The main limitation is that the cost of the optimization process of QPD will be expensive for very high-dimensional models, e.g., computer vision models, often with hundreds of thousands of dimensions.
Consequently, a potential remedy is to find the correlation among noise dimensions and eliminate these to reduce the optimization complexity.
This also gives rise to an urgent call for investigations on the development of techniques for model extraction mitigation, presenting potential opportunities to the research community.

Although MDP can defend against model extraction attacks such as QPD, it mainly relies on slowing down the attack speed to make QPD reach the desired extraction level far beyond the normal number of queries. This actually does not eliminate the existence of this kind of attack.
Correspondingly, the performance of MDP in the high-dimensional model also decreased.
It is also worth studying how to extend MDP to defend against other kinds of subversion attacks, including evasion and poisoning.
In the following research, we will transition from model extraction to function extraction. After all, function protection should be the core task of model attacks.
\\

\noindent\textbf{Extension.}
We plan to apply the QPD attack to steal the information of DNNs and MDP to perform detection and defense due to DNNs attract plenty of attention.
The DNN model is significantly more complex in depth and hyper-parameter settings than the LR and NN models currently studied. 
As we all know, when faced with the model extraction attack, the simpler the model, the more difficult it is to defend.
Furthermore,
QPD is considered in the functional direction of the model. As long as the target provides services, it is possible to extract a similar function model from the query results.
MDP is considered from the dataset behind the model, regardless of the type of model.
Therefore, we believe that we only need to fine-tune our mechanism for a specific model, QPD and MDP can achieve good results.

The other is defending against collusion attacks. We think that this type of attack needs to be designed and considered more from the service mechanism aspect, which is not the main problem addressed in this paper. 
Nevertheless, existing work claims to measure the knowledge gained by colluding adversaries, and our proposed Monitor can do a better job along this line. We believe that this issue is still worth re-examining.

\section{Conclusion} \label{Conclusion}
In this work, we develop a novel, adaptive query-flooding parameter duplication attack. By only accessing the public APIs, an attacker can infer the private machine learning models protected by the state-of-the-art differential privacy and monitoring mechanisms via QPD.
After that, a defensive solution against the attack is proposed.
We design a new real-time model extraction status assessment scheme called \textit{Monitor} first.
By quantifying the loss of privacy of training data from such attacks,  \textit{Monitor} reflects the current extraction status of the model in a more accurate and timely manner.
Then, we introduce the monitoring-based differential privacy mechanism to protect the model. Unlike the previous work, the new defense dynamically adjusts the amount of noise added in the response with the support of a new adaptive differential privacy budget allocation method and effectively defends the QPD attack.
In addition, 
experiments have thoroughly evaluated the proposed schemes.

As we all know, the attack is not the ultimate goal.
It is still worth studying similar attack techniques, which allows us to understand the related vulnerabilities better and provide insightful guidance for corresponding defense strategy development.



\bibliographystyle{IEEEtran}
\bibliography{IEEEabrv,HDGM}

\end{document}